\documentclass[]{spie}  

\usepackage{amsmath,amsfonts,amssymb}
\usepackage{graphicx}
\usepackage[colorlinks=true, allcolors=blue]{hyperref}
\usepackage[dvipsnames]{xcolor}
\usepackage{MnSymbol}
\usepackage{subcaption}

\title{Adapting the pyramid wavefront sensor for pupil fragmentation of the ELT class telescopes}

\author[a,b,d]{Nicolas Levraud}
\author[a,b]{Vincent Chambouleyron}
\author[c]{Olivier Fauvarque}
\author[a,b]{Mahawa Cisse}
\author[a,b]{Jean-François Sauvage}
\author[b]{Benoît Neichel}
\author[e]{Charlotte Bond}
\author[d]{Enrico Pinna}
\author[d]{Simone Esposito}
\author[e]{Noah Schwartz}
\author[a,b]{Thierry Fusco}

\affil[a]{DOTA, ONERA, Université Paris Saclay, F-91123 Palaiseau, France}
\affil[b]{Aix Marseille Univ, CNRS, CNES, LAM, Marseille, France}
\affil[c]{IFREMER, Laboratoire Detection, Capteurs et Mesures (LDCM), Centre Bretagne, ZI de la Pointe du Diable, CS 10070, 29280, Plouzane, France}
\affil[d]{INAF - Osservatorio Astrofisico di Arcetri}
\affil[e]{UK Astronomy Technology Centre, Blackford Hill, Edinburgh EH9 3HJ, United Kingdom}

\authorinfo{Further author information: (Send correspondence to N.Levraud)\\N.Levraud : E-mail: nicolas.levraud@lam.fr, Telephone: +33 788 33 01 02\\  }

\begin{document} 
\maketitle

\begin{abstract}
The next generation of Extremely Large Telescope (24 to 39m diameter) will suffer from the so-called "pupil fragmentation" problem. Due to their pupil shape complexity (segmentation, large spiders ...), some differential pistons may appear between some isolated part of the full pupil during the observations.
Although classical AO system will be able to correct for turbulence effects, they will be blind to this specific telescope induced perturbations. Hence, such differential piston, a.k.a petal modes, will prevent to reach the diffraction limit of the telescope and ultimately will represent the main limitation of AO-assisted observation with an ELT. In this work we analyse the spatial structure of these petal modes and how it affects the ability of a Pyramid Wavefront sensor to sense them. Then we propose a variation around the classical Pyramid concept for increasing the WFS sensitivity to this particular modes. Nevertheless, We show that one single WFS can not  accurately and simultaneously measure turbulence and petal modes. We propose a double path wavefront sensor scheme to solve this problem. We show that such a scheme,associated to a spatial filtering of residual turbulence in the second WFS path dedicated to petal mode sensing, allows to fully measure and correct for both turbulence and fragmentation effects and will eventually restore the full capability and spatial resolution of the future ELT.


\end{abstract}


\section {Introduction }
\label{sec:intro} 
Adaptive Optics (AO) is necessary today to take the full advantage of the size of modern telescopes. Without the compensation of atmospheric turbulence by AO, the only advantage of bigger telescope is the larger surface of light collection, the angular resolution being the one of a telescope of size $r_0$, the Fried parameter. The improvement in resolution allowed by AO is crucial for applications such as exoplanet direct imaging. These systems are composed of 3 parts: a Wavefront Sensor (WFS) which measures the effect of atmospheric turbulence on the incoming wavefront, a Real Time Computer [RTC] which computes, in real time and at high frequency (typically a few hundreds to a few thousands of Hz depending on the final requirements), the correction to be applied, and eventually a Deformable Mirror (DM) which applies these corrections to the wavefront. One of the fundamental limits for any AO system is the quality of the wavefront measurements produced by the WFS. In particular the performance of a WFS can be defined with respect to two criteria : its sensitivity (the ability to measure a given perturbation with a given accuracy using a limited number of photon) and its dynamic range (the maximum amplitude of the incoming wavefront for which the WFS signal can be assumed to be linear). A wavefront sensor sensitive to a given phase mode will give a strong signal for a very small variation of the mode. A sensor with a high dynamic for a mode is able to measure high amplitude of this mode with while keeping a good accuracy. The classical wavefront sensor used for AO, the Shack-Hartmann wavefront sensor (SHWFS) by its nature gets less sensitive as the number of sub-apertures increases.A larger primary mirror needs more mode to be controlled to reach its diffraction limit, hence the SHWFS becomes less and less efficient for larger telescope. In order to partially mitigate this aspect, a more recent WFS concept, the Pyramid Wavefront Sensor (PyWFS)\cite{schwartzSensingControlSegmented2017} has been chosen for the next generation of Extremely Large Telescope. Such a Fourier Filtered based WaveFront sensor concept\cite{fauvarqueGeneralFormalismFourierbased2016} can (partially or fully depending on the sensor design and the operational conditions) benefit from the "Full aperture gain" and thus overtakes the SH sensitivity (by at least a few magnitudes). 
In parallel, these telescopes due to their size have a new difficulty built-in : the pupil fragmentation. Due to the large gap in some parts of the pupil (for instance generated by large spider shadows projected onto M1 surface), the continuity of the wavefront between these "pupil fragments" is not guaranteed. This can creates some so-called petal modes \cite{schwartzSensingControlSegmented2017} where the AO system corrects well the wavefront inside a pupil fragment but is unable to ensure a good phase continuity between segments. Hence, a random distribution of pistons could appear between each fragment. This pistons can be significantly larger than one wavefront senor wavelength and thus they will ultimately degrade the final image quality. Specific solutions are to be measured and controlled.

The aim of this work is to see in which conditions the PyWFS fails to measure these petal mode and to propose solutions to overcome this current limitation. In the first part of this article, we will study the current limitation of classical PyWFS to pupil fragmentation effects. We will identify the physical reasons of the sensitivity loss associated to the Pupil fragmented modes. In the second part we will propose an alternative design allowing to overcome the previously identified limitations. We will analyse the performance of this new spatially filtered PyWFS and demonstrate its capacity to produce accurate and robust measurements of the pupil fragmented modes even under strong turbulence regime.


\section {Measurement of pupil fragmentation effects with a Pyramid wavefront sensor}

\subsection {The petal modes}

\begin{figure}[htbp]
    \centering
    \begin{subfigure}{0.2\textwidth}
    \includegraphics[width=0.9\linewidth]{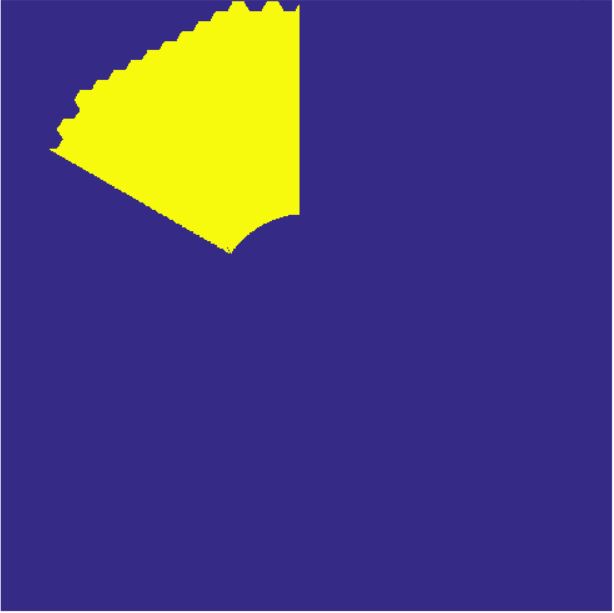}
    \end{subfigure}
    \begin{subfigure}{0.2\textwidth}
    \includegraphics[width=0.9\linewidth]{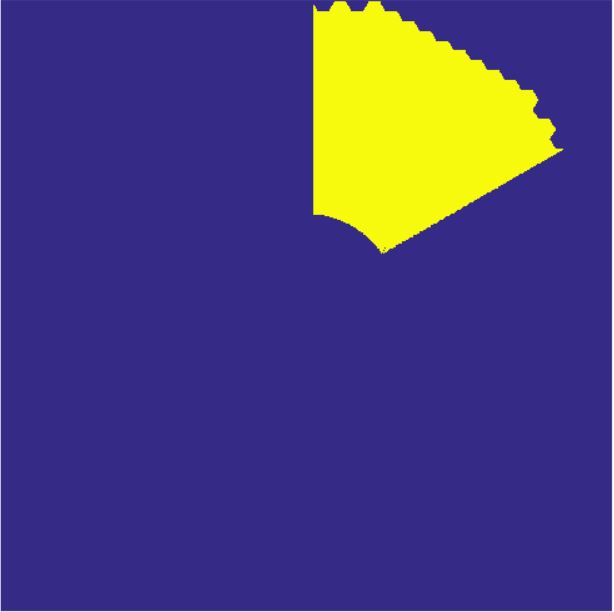}
    \end{subfigure}
    \begin{subfigure}{0.2\textwidth}
    \includegraphics[width=0.9\linewidth]{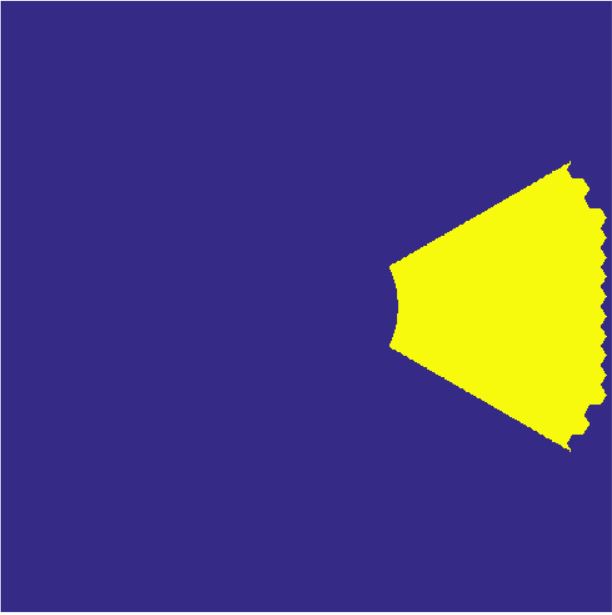}
    \end{subfigure}
    \begin{subfigure}{0.2\textwidth}
    \includegraphics[width=0.9\linewidth]{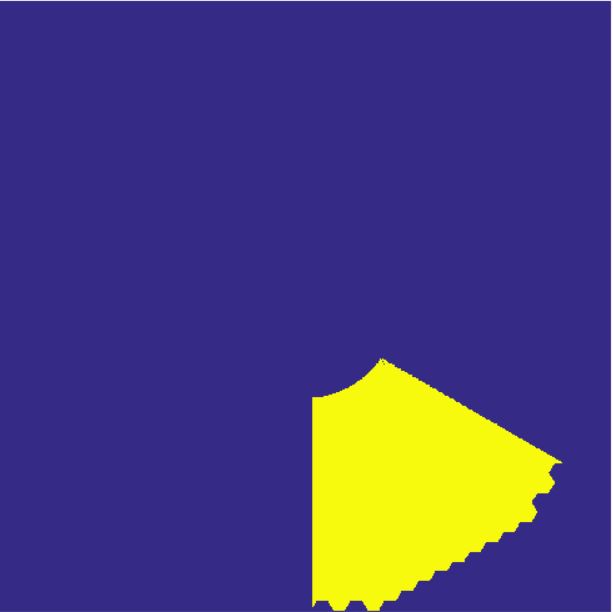}
    \end{subfigure}    
    \begin{subfigure}{0.2\textwidth}
    \includegraphics[width=0.9\linewidth]{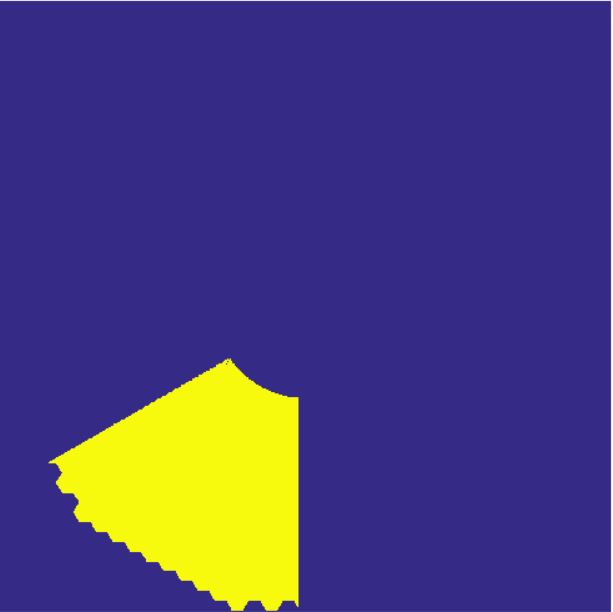}
    \end{subfigure}
    \begin{subfigure}{0.2\textwidth}
    \includegraphics[width=0.9\linewidth]{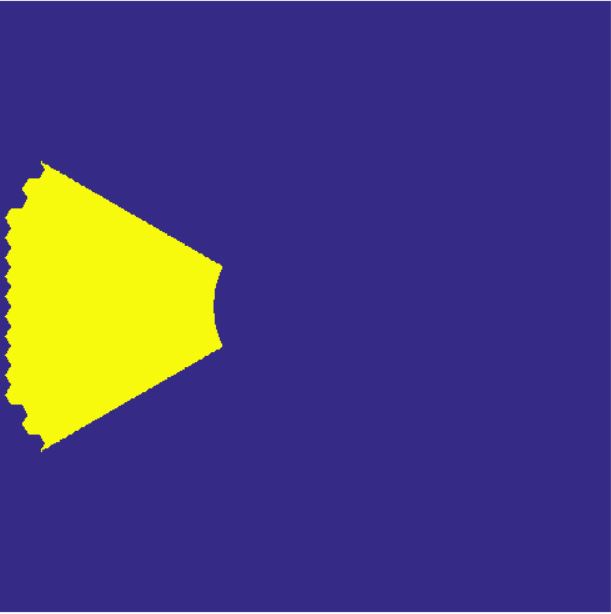}
    \end{subfigure}
    \caption{ELT fragments}
    \label{fig:my_label}
\end{figure}

The pupil fragmented effects are nothing but  differential pistons produced either by the AO loop itself (propagation of unseen modes in the loop process) or by local phase defects produced by the telescope itself combined with pupil discontinuities generated by its design (produced by the shadow of the large spiders  supporting M2 for instance) in the pupil plane. As a results, a classical AO system will perform an efficient wavefront measurement and correction inside each pupil fragment (delimited by two spider shadows) but can not measure pupil discontinuities (i.e. the differential piston between to segments) especially when spider width is typically larger the the fried parameter at the wavefront sensing wavelength. In the ELT case, spiders of 50cm will create six fragments in the pupil (see Fig.\ref{fig:my_label}. 
An ELT pupil fragment (or petal) $p_{ij}$ is defined as follow 
\begin{equation}
    p_{ij}=a_i-a_j
\end{equation}
with
\begin{equation}
    a_i = \frac{I}{S_{fragment}} \int_{x,y \in fragment} \phi(x,y) dx dy\label{eq:petal}
\end{equation}

We can therefore defined the Pupil Fragmented or Petal modes which will be nothing but a linear the combination of these fragments\cite{bertrou-cantouValidationComposantsClefs2018}.

The purpose of this study is to better understand the coupling of these petal modes with classical Pyramid-based AO systems and to propose innovative approaches to mitigate their deleterious effects on the final AO loop performance.  To do so, we will first study the sensitivity of the classical Pyramid (with or with modulation) to these petal modes only (without any turbulence) and we will propose some modification of the Pyramid concept (adapted to the complex pupil geometry) that will allow to increase the final WFS sensitivity to this mode. Then we will couple petal mode and turbulence residue and propose a 2 stages modified Pyramid concept that will be able to simultaneously measure both classical turbulence and specific petal modes.

For the sak eof simplicity and clarity we 
will here work with a simplfied "toymodel" of a fragmented pupil (limited to 2 fragmented pupils). Such a simplifed toolls will allow us to better understand the sensitivity of a PyWFS to the petal modes and to thoroughly study the main parameters (spider size, pyramid modulation) that affect the performance of the sensor.

\begin{figure}[ht]
    \centering
    \includegraphics[width=12cm]{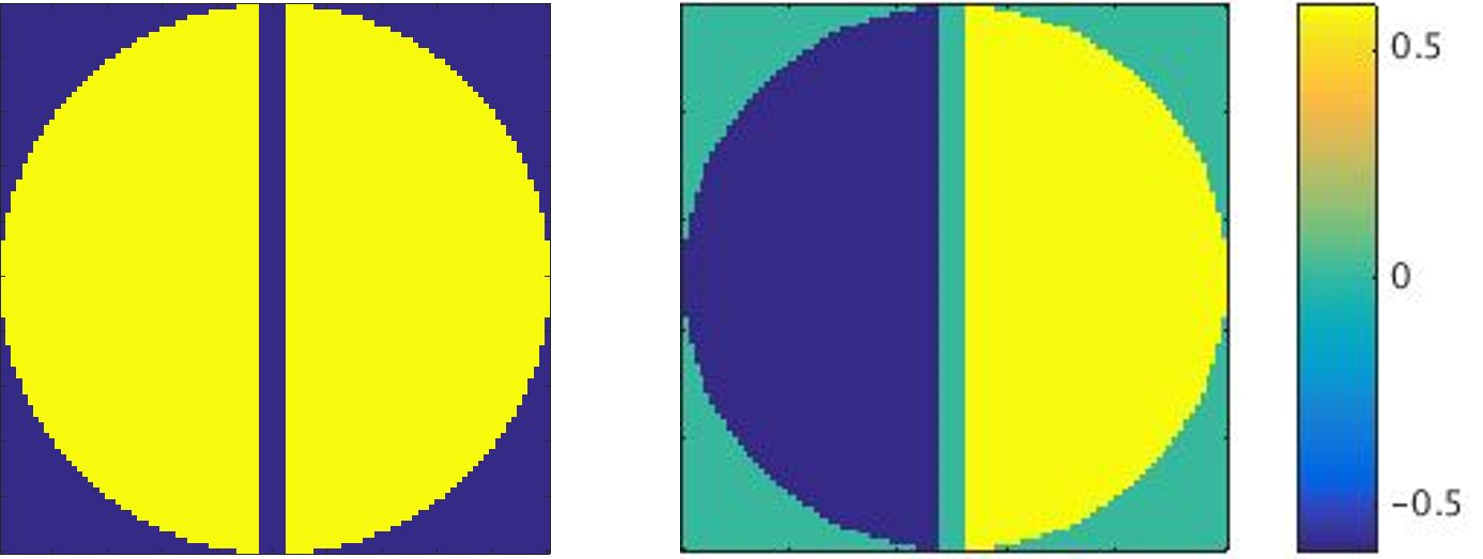}
    \caption{Toymodel pupil and petal mode $P_{Toy}$}
    \label{fig:toymodel}
\end{figure}

 In the following we will consider a 10m diameter telescope cut in half by a "spider" of a variable size (from 0 to 1m). This configuration produce a pupil shape and and one single petal mode $P_{Toy}$ (to which we associated a given amplitude $p_{Toy}$)
as defined in Eq.~\ref{eq:petal} and shown in Fig.~\ref{fig:toymodel}. 
 The aim of this toymodel is to simplify the problem to only one petal to consider and still keep representatives conditions where the petal mode will appear in an AO loop. The use of only one petal modes also greatly reduces the computational burden of the simulation. We create an end to end simulation using the OOMAO  Toolbox\cite{conanObjectorientedMatlabAdaptive2014}
to simulate the toymodel and the pyramid signal. This toolbox is also used for the AO loop simulations later in this paper.  

\subsection {Classical pyramid response}
The very first parameter we study is the response of the PyWFS to a petal mode without any other perturbation. Such a response will  give us the ultimate sensitivity of the WFS to this particular mode for various spider width and PyrWFS configuration (modulation radius).

\begin{figure}[ht]
    \centering
    \includegraphics[width=12cm]{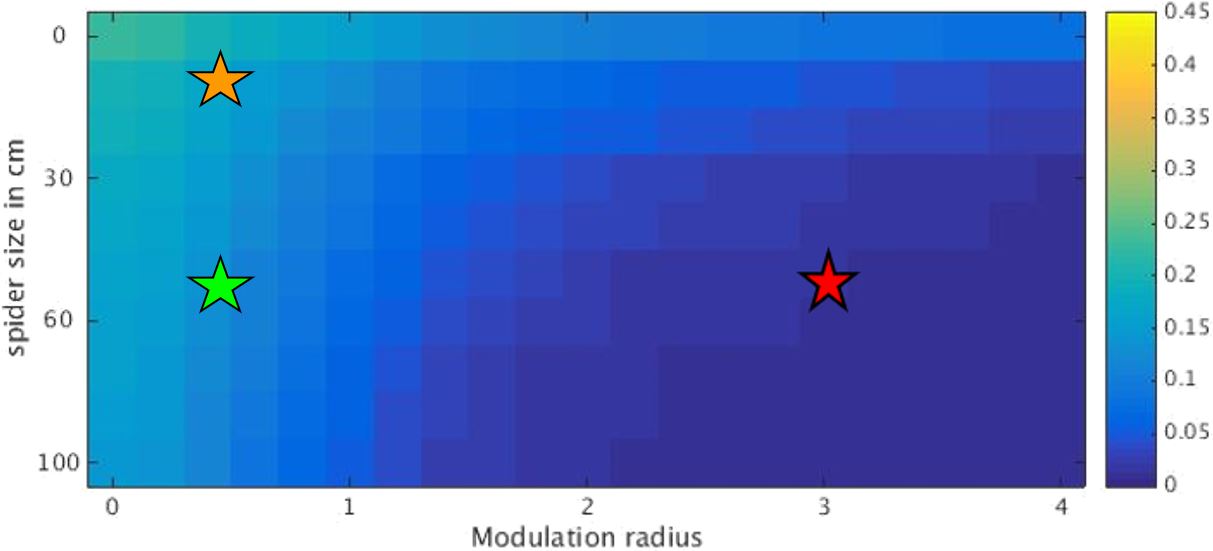}
    \caption{Sensitivity computed for petal mode and different pupil and modulation size. \textcolor{red}{\huge$\filledstar$} SCAO system of Harmoni on ELT \textcolor{orange}{\huge$\filledstar$}  Sphere + \textcolor{green}{\huge$\filledstar$} future ExAO system on ELT }  
    \label{fig:S4sidedpyr}
\end{figure}

To do so, we use the method developed in  \cite{fauvarqueOptimisationAnalyseursFront2017}.  
If we call $I(\phi) $ the intensity recorded on the PyWFS detector, we compute the response to a mode $\delta I(\phi)$ as :
\begin{equation}
    \delta I(\phi) = \frac{I(\epsilon*\phi)- I(-\epsilon*\phi)}{2*\epsilon*N_{ph}}
    \label{eq:Ivar}
\end{equation}

\begin{equation}
    S(\phi) = ||\delta I (\phi)|| ^2
\end{equation} 

with F = total flux on the detector, $\epsilon$ the amplitude of the mode used, $\phi$ the phase mode.$\epsilon$ needs to be low enough to be in the linear regime.  S is the sensitivity to the phase $\phi$. It corresponds to the variation of signal for a small phase variation. Generally speaking, for the PyWFS the signal is nothing but the intensity measured on the detector. Using a small variation of phase guarantees we stay with the linear regime of the sensor. The more sensitive to a mode our sensor is, the more resilient to noise it is so the least light can be used by the WFS. As was shown in \cite{chambouleyronOptimisationAnalyseSurface}, this sensitivity computation is valid for readout noise.
\begin{figure}[ht]
    \centering
    \includegraphics[width=12cm]{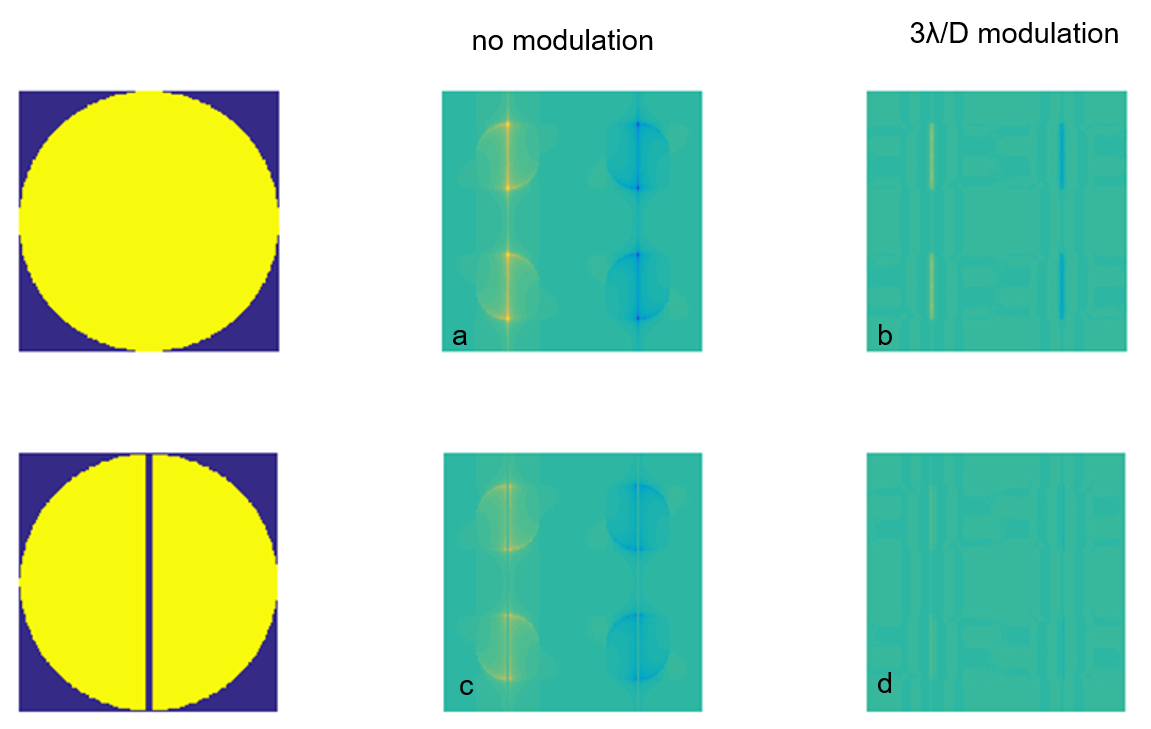}
    \caption{$\delta$I(petal) for different pyramids : a) no modulation no spider pupil b) modulation
3$\lambda$/D no spider pupil c) no modulation but spider d) 3$\lambda$/D modulation with spider}
    \label{fig:deltaIclassicalpyr}
\end{figure}
In Fig.~\ref{fig:deltaIclassicalpyr} we simulate the pyramid response (the Intensity variations defined above Eq\ref{eq:Ivar})  case with and without spider. a and b are without spider, c and d with spiders of 50cm widt. a and c without modulation, b and d with modulation. The most important result is the modification of the signal when using modulation. It is common to say that a modulated pyramid is a slope sensor while a non modulated one is a phase sensor \cite{verinaudNatureMeasurementsProvided2004}. We see this behaviour in this case with the signal of petal mode being distributed in the whole pupil for the non modulated pyramid while it is concentrated at the edge of the discontinuity for a modulated pyramid. In the presence of a spider that means that the photons which carry the information are hidden by the spider and do not reach the detector. Hence, in the case of both modulation and presence of spider the WFS response to a petal mode signal is close to 0. 
We then explore (Fig.~\ref{fig:S4sidedpyr}) the space parameter of spider width, modulation, and compute the sensitivity to the petal mode for various combinations of those. 
This ``Petal sensitivity map''  shows that the pyramid becomes less and less to petal mode with the increase of modulation and spider width. In an HARMONI SCAO configuration for example, where the modulation is set to 3 Lambda/D and the spiders width 50cm, the sensitivity is smaller than 0.05 and only reached 0.2  for a non-modulated pyramid without any spider (2 being the Cramer Rào upper bound for any Fourier-Filter WFS). The factor 10 between the Cramer Rào limit and the non-modulated pyramid without spider (it can go up to a few hundreds when modulation and spiders are considered) definitely shows the huge potential gain we could expect by modifying the type of mask and the modulation strategy in the WFS. 
This is the purpose of the next section. 

\subsection {Asymmetric Pyramid}
\begin{figure}[ht]
    \centering
    \includegraphics[width=16cm]{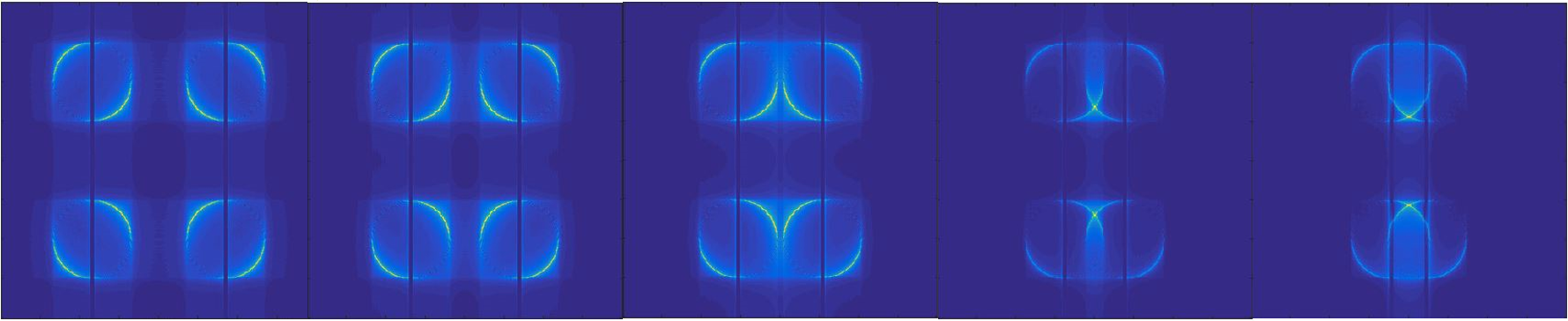}
    \caption{Change in reference intensity when reducing the angle in the X direction of a 4 sided pyramid}  
    \label{fig:4sidedtoasympyr}
\end{figure}

\begin{figure}[ht]
    \centering
    \includegraphics[width=16cm]{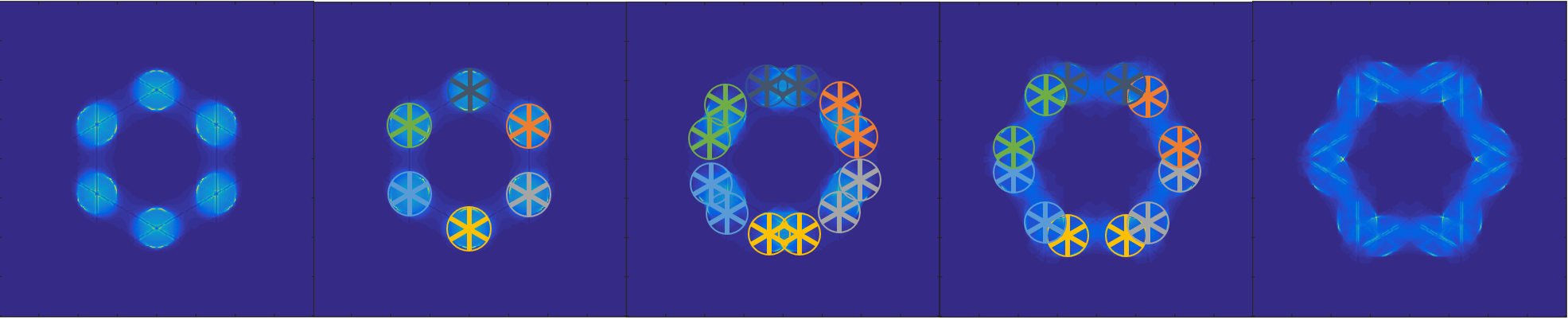}
    \caption{Change in intensity following the ELT subpupil when asymetrising the 6 sided pyramid to adapt to ELT petals. }  
    \label{fig:Star pyr}
\end{figure}

Following the idea proposed by O. Fauvarque a couple of years ago we propose a variation around the Flatten Pyramid theme\cite{fauvarqueVariationPyramidTheme2015} . The idea is to optical combine the signal coming from different fragmented part of the pupil BEFORE the detection.
The way for that was to take advantages of the symmetries of the system and force fragments to interfere with each other.  
Practically speaking, we have taken the classical pyramid and increase the pyramid angle in one direction, breaking the X/Y symmetry as shown in the Fig.~\ref{fig:4sidedtoasympyr} where the X-direction angle is modified. 
The parameter of the asymmetry angle need to be optimised and the idea can be used for various kind of pupils with symmetric fragments.  Another example is shown in Fig.~\ref{fig:Star pyr}  for the ELT pupil. 

\begin{figure}[ht]
    \centering
    \includegraphics[width=12cm]{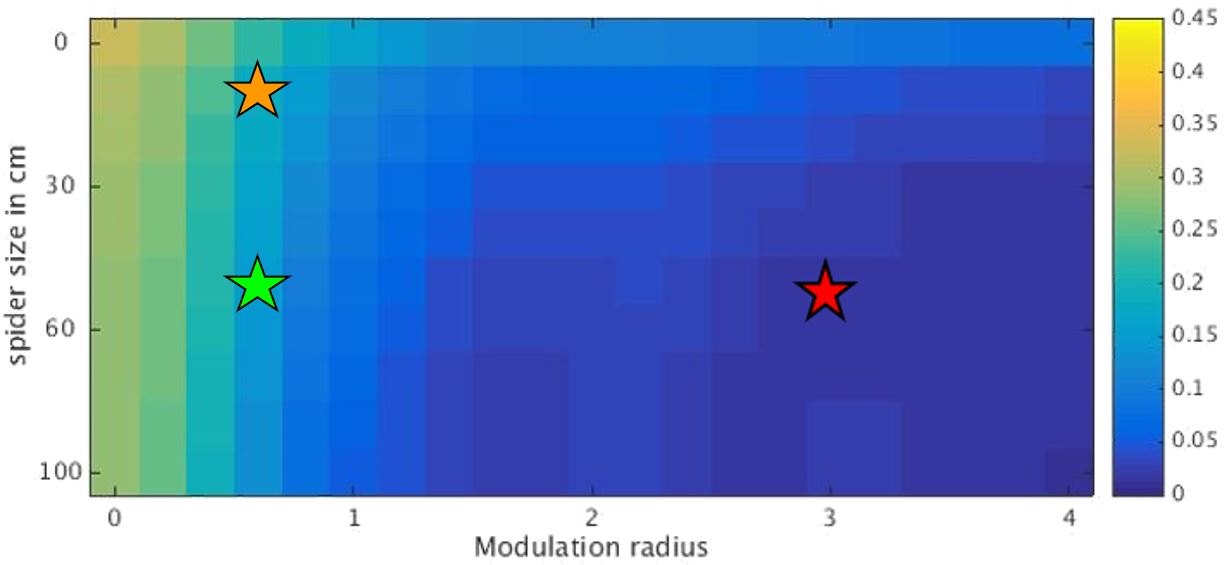}
    \caption{Sensitivity computed for petal mode and different pupil and modulation size. \textcolor{red}{\huge$\filledstar$} SCAO system of HARMONI on ELT \textcolor{orange}{\huge$\filledstar$}  Sphere + \textcolor{green}{\huge$\filledstar$} future ExAO system on ELT }  
    \label{fig:Sasympyr}
\end{figure}

The sensitivity map for the asymmetric pyramid has been computed as well Fig.~\ref{fig:Sasympyr} 
This new asymetric configuration significantly increase (typically by a factor two to three)  the signal in response to the petal mode but after testing the modulated case we see the same behaviour with signal reduced when the modulation radius augments. 

 In the presence of spider and modulation the asymmetric PyWFS remains barely sensitive (though 2 or 3 times better than the classical case) to petal mode. 
 Indeed, when we look at the power spectral density of a petal mode, most of its energy is concentrated in the low frequencies. The drop of sensitivity with modulation is therefore completely normal and will be present whatever the focal plan filter. Only non barely modulated sensor will allow to recover petal mode signal. The drawback of such a configuration beeing a very small dynamic range. 
 This has two consequences : 
 \begin{itemize}
 \item only small petal mode amplitude could be measured 
 \item non modulated sensor will NOT be able to work under classical turbulence conditions at least at visible wavelengths. 
 \end{itemize}
For classical AO system (linear reconstructor using Matrix Vector Multiplication), the direct consequence is the need of 2 separated WFS in the AO system : one for turbulence measurement, the other dedicated to  to the petal mode (hereafter called petalometer) in the rest of this paper.  Alternatively there are studies to find reconstructors able to compensate for the non linearity. For now we stay with a linear  integrator for control and modulation is necessary.
The next section will detail this specific point.

\section{ Dealing with turbulence}

\subsection{ 2 paths AO system}
Following the previous results a 2 path AO system has been simulated end to end to test our understanding of the system. It is composed by 
\begin{itemize}
\item a main modulated PyrWFS whose purpose is to measure the turbulent phase. 
\item a non modulated Pyramid dedicated to the petaling mode measurement only
\end{itemize}

\begin{figure}[ht]
    \centering
    \includegraphics[width=12cm]{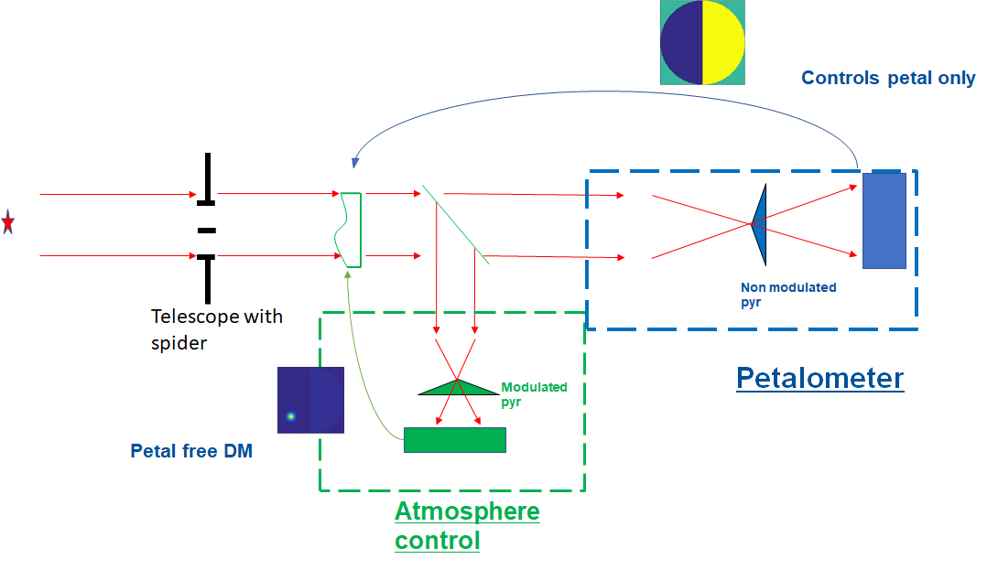}
    \caption{2 Path AO system simulated}  
    \label{fig:2Pathsystem}
\end{figure}

The system is using 2 WFS with separated controllers. The only common elements they have is the DM for which they control different modes. To make sure the 2nd path sensor would not measure phase introduced by the 1st path, all modes controlled by the 1st path sensor had their petal filtered out. First we need to compute the projection of each of the DM actuators $\phi_{act}$ on the Petal mode : 
\begin{equation}
    \hat{p}=<\phi_{act}|P_{toy}>
\end{equation}
\begin{equation}
     \phi_{petalfree} = \phi_{act} - \hat{p}P_{toy}
\end{equation}

The conditions for the simulation are as follow : 
\begin{table}[ht]
\caption{Conditions of AO loop simulation.} 
\label{tab:SimConditions}
\begin{center}       
\begin{tabular}{|l|l|}
\hline
\rule[-1ex]{0pt}{3.5ex} Deformable Mirror & 20$\times$20 gaussian influence function + petal mode  \\
\hline
\rule[-1ex]{0pt}{3.5ex}  Telescope Pupil parameter & 10m telescope, 50cm spider  \\
\hline
\rule[-1ex]{0pt}{3.5ex}  $r_0$ & 17cm@550nm  \\
\hline
\rule[-1ex]{0pt}{3.5ex}  Wind speed & 5m/s  \\
\hline
\rule[-1ex]{0pt}{3.5ex}  Noise & no noise  \\
\hline
\rule[-1ex]{0pt}{3.5ex}  PyWFS sensing wavelength & 550nm monochromatic simulation  \\
\hline
\rule[-1ex]{0pt}{3.5ex}  PyWFS Subpupil number & 100 subpupil  \\
\hline
\rule[-1ex]{0pt}{3.5ex}  AO Loop frequency & 500Hz  \\
\hline
\end{tabular}
\end{center}
\end{table} 

Two parameters are tracked to see the quality of wavefront correction : the wavefront RMS and the petal mode present in the phase. Since we consider monochromatic light the petal mode is defined modulo $\lambda$. The petal at frame 0 is subtracted to every frame to start from a phased configuration. The DM correction computation uses a simple Matrix integrator with no optical gain control. The computation uses the full frame reduced intensity method laid out in \cite{fauvarqueOptimisationAnalyseursFront2017}.

\begin{figure}[ht]
    \centering

   \begin{subfigure}{0.45\textwidth}
\includegraphics[width=0.9\linewidth]{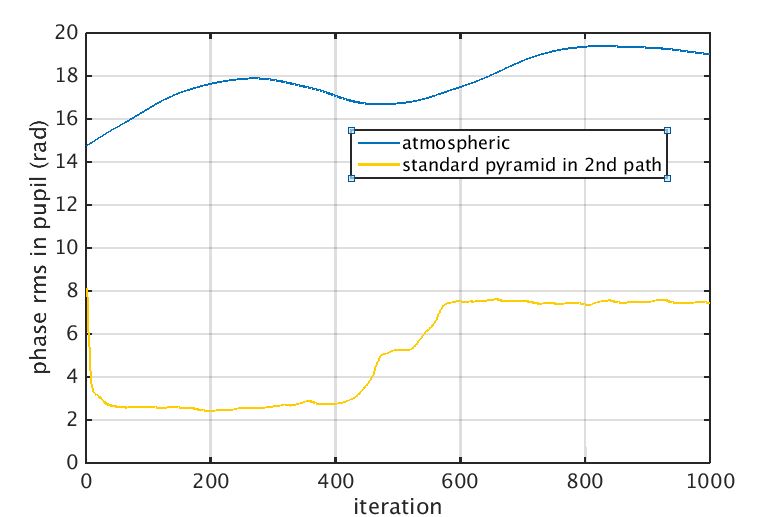} 
\caption{RMS of phase residual}
\label{fig:subim1}
\end{subfigure}
\begin{subfigure}{0.45\textwidth}
\includegraphics[width=0.9\linewidth ]{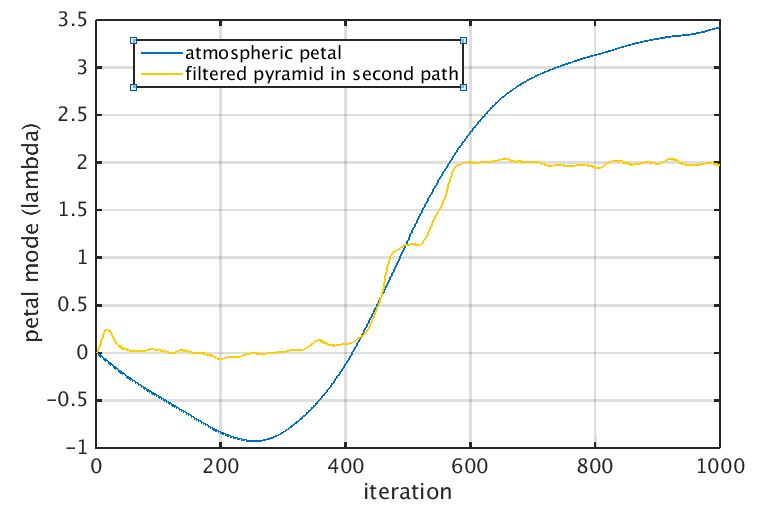}
\caption{Residual projected on petal mode}
\label{fig:subim2}
\end{subfigure}
    \caption{AO loop result for 2 path sensors}
    \label{fig:Result2pathnofilter}
\end{figure}

We can see on Figure \ref{fig:Result2pathnofilter} While the petal is controlled by the 2nd path sensor it is not resilient enough.  $\lambda$ jumps appear from time to time. This means that despite being used in presence of already compensated phase, the non-modulated pyramid used as petalometer is not working in its linear regime. A additional way to reduce the impact of residual turbulence on the petal control is needed.

\subsection{Filtering out the residual turbulence in the second stage}
Practically speaking there are only four different ways for reducing the impact of phase residuals in the second path WFS, they are detailed below with their potential drawbacks : 
\begin{table}[ht]
\caption{Ways to reduce impact of atmospheric residuals} 
\label{tab:IncreasePetalVSresidual}
\begin{center}       
\begin{tabular}{{|}p{0.4\textwidth}{|}p{0.4\textwidth}{|}}
\hline
\rule[-1ex]{0pt}{3.5ex} Increase level of 1st path correction  &  \textcolor{red}{increase of AO complexity and reduction of the limit magnitude}  \\
\hline
\rule[-1ex]{0pt}{3.5ex} Use longer wavelength  & \textcolor{red}{System more complex and some science photon potentially stolen by WFS}  \\
\hline
\rule[-1ex]{0pt}{3.5ex}Temporal filtering  (increase of 2nd stage integration time)  &  \textcolor{red}{Limit Petal mode correction bandwidth and thus its efficiency}  \\
\hline
\rule[-1ex]{0pt}{3.5ex}  Spatial filtering & \textcolor{Green}{Compatible with SCAO and visible WFS}   \\
\hline

\end{tabular}
\end{center}
\end{table} 

When looking at the Power Spectral Density (PSD) of the petal mode VS residual phase (assuming the same total RMS error), the spatial distribution of the energy for the two cases is very different. While most of the energy of the petal mode is located on the lowest spatial frequencies, the turbulence residuals has its energy mainly distributed after the AO cut-off. We can therefore identify to separate regimes:
\begin{itemize}
    \item a low frequency regime (typically up to a few cycles per pupil) where the Petal PSD is clearly the dominant signal
    \item a medium to high frequency regime fully dominated by turbulence residuals
\end{itemize}
By using a spatial filter it should therefore be possible to cancel out most atmospheric residuals while keeping petal untouched.
\begin{figure}[ht]
    \centering
    \includegraphics[width=16cm]{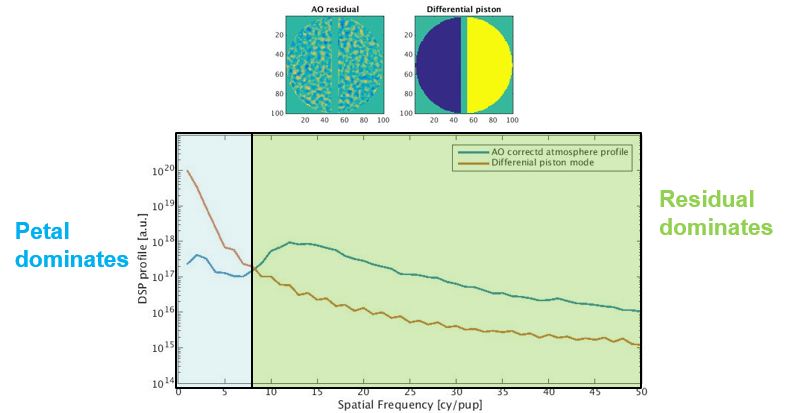}
    \caption{PSD of petal mode VS PSD of atmospheric residuals }  
    \label{fig:PSDpetalVSresidual}
\end{figure}
 To check our hypothesis we take petal and residual turbulent phase in a pupil plane, pass them through spatial filters of various size and propagate them again to a pupil plane.
\begin{figure}[ht]
    \centering
    \includegraphics[width=16cm]{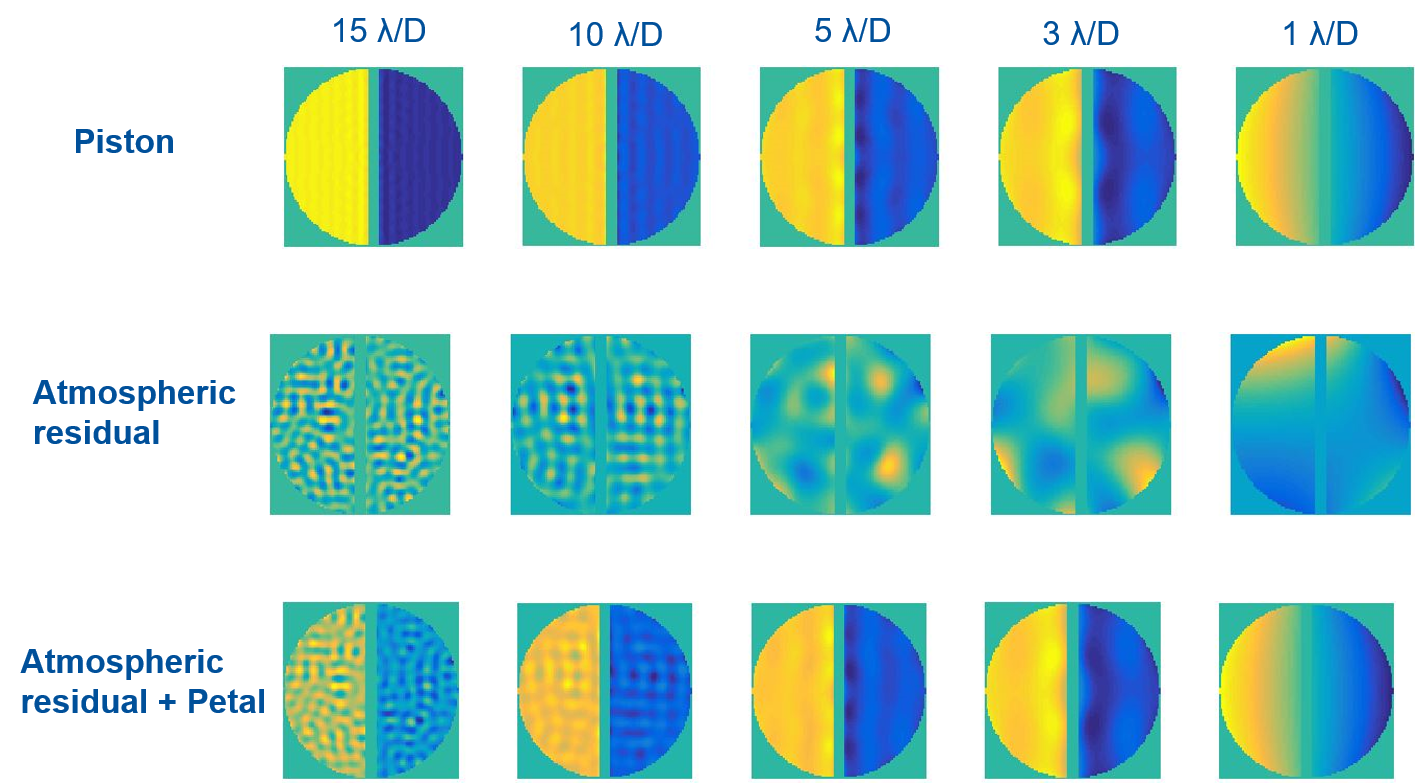}
    \caption{Phase in pupil change after a focal plane  }  
    \label{fig:Phasefiltereffect}
\end{figure}
Fig.~\ref{fig:Phasefiltereffect} shows that  Petal mode is very resilient to focal plane filtering and keeps its characteristic shape till a very small filter size (typically smaller than 3 $\lambda/D$. Meanwhile the residual turbulence is very rapidly smooth out. Although an optimal size of the pinhole can certainly be found using quantitative criteria, a simple qualitative analysis of the results clearly shows that 5 $\lambda/D$ brings a very good compromise between residual turbulence reduction and petaling signal preservation. 

\subsection{Filtered non-modulated PyWFS as petalometer}
\begin{figure}[ht]
   \centering
   \begin{subfigure}{0.23\textwidth}
\includegraphics[width=0.9\linewidth]{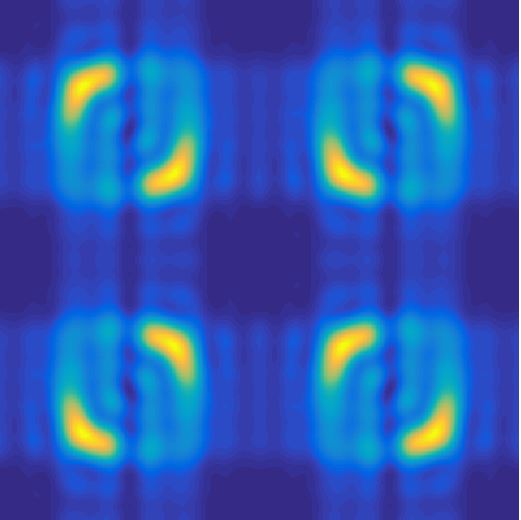} 
\caption{ Reference Intensity for 4$\lambda$/D filtering}
\label{fig:Iref4L/D}
\end{subfigure}
   \begin{subfigure}{0.23\textwidth}
\includegraphics[width=0.9\linewidth]{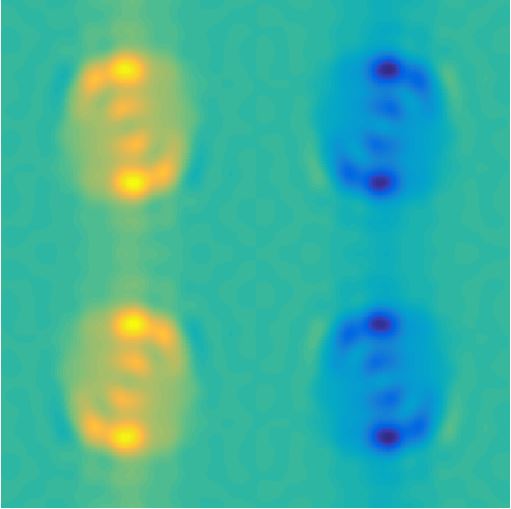} 
\caption{ $\delta I(petal)$ for 4$\lambda$/D filtering}
\label{fig:deltaI4L/D}
\end{subfigure}
   \begin{subfigure}{0.23\textwidth}
\includegraphics[width=0.9\linewidth]{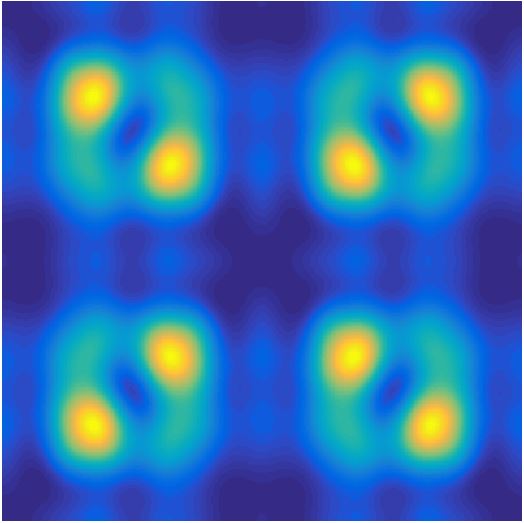} 
\caption{ Reference Intensity for 2$\lambda$/D filtering}
\label{fig:Iref2L/D}
\end{subfigure}
\begin{subfigure}{0.23\textwidth}
\includegraphics[width=0.9\linewidth ]{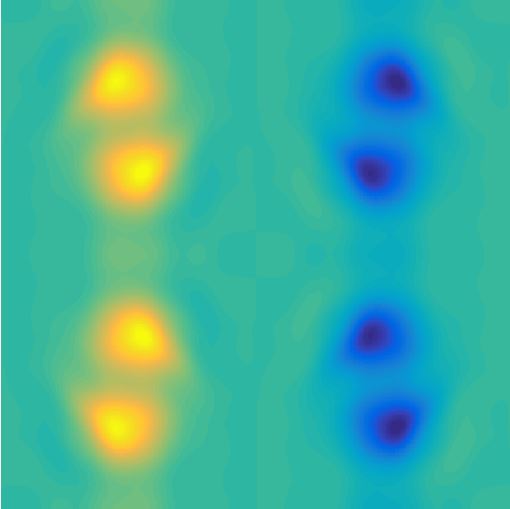}
\caption{ $\delta I(petal)$ for 2$\lambda$/D filtering}
\label{fig:deltaI2L/D}
\end{subfigure}
\caption{Filtered pyramid Intensity}
\label{fig:Pyramidfiltereffect}
\end{figure}

Following the previous results a new AO system is simulated with a focal plane filter added on top of the PyWFS in the petalometer branch. This focal plane filter first changes both the reference intensity and the interaction matrix for the PyWFS cf Fig.~\ref{fig:Pyramidfiltereffect}

\begin{figure}[ht]
\centering
\begin{subfigure}{0.45\textwidth}
\includegraphics[width=0.9\linewidth]{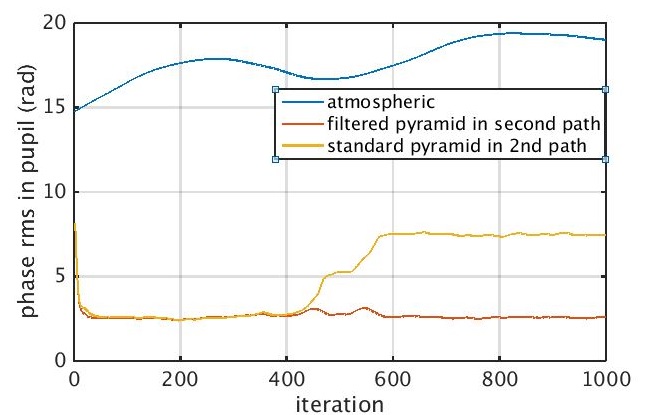} 
\caption{RMS of phase residual}
\label{fig:rms_filtered _effect}
\end{subfigure}
\begin{subfigure}{0.45\textwidth}
\includegraphics[width=0.9\linewidth ]{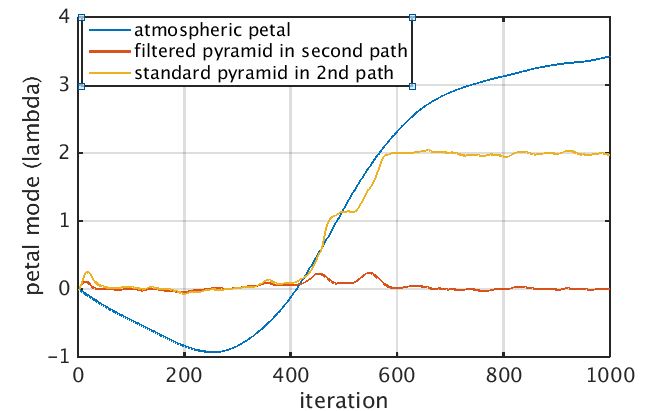}
\caption{Residual projected on petal mode}
\label{fig:petal_filtered_effect}
\end{subfigure}
\caption{AO loop result with filtered pyramid as petalometer}
\label{fig:Result2pathfilter}
\end{figure}

Even though the interaction matrix changes the process stay the same. Here are the AO loop results : Fig.~\ref{fig:Result2pathfilter} with a  2$\lambda/D$ filter compared to the previous simulation. The simulation is done with exactly the same parameters (and atmospheric conditions) except the filter added on the same plane as the PyWFS.

As expected, the filtered non-modulated Pyramid allows to solve the non-linearity issues encountered by the classical non-modulated case. The AO loop remains extremely stable without any $\lambda$ jump without any loss in terms of turbulence correction. 

\section{CONCLUSION and perspectives}

Measuring and controlling pupil fragmentation effects (petaling, low wind effect) is one of the most important challenge for the future AO systems of the ELT. The very peculiar spatial structure of the differential piston produced by the pupil fragmentation makes it impossible to measure with the wavefront sensor originally used to measure for turbulence perturbation. A dedicated sensor (petalometer) must be considered and optimize for the specific measurement of the petaling mode. 
In order to have the most possible resilient sensor, residual turbulence must be filtered out at the intrance of the petalomete. Among the various way to reduce the impact of atmospheric residuals the spatial filtering is probably one of the most promising one. In this paper We have shown the efficiency of the coupling of a spatially filtered non-modulated Pyramid for measuring the differential piston. Note that the spatially filtered stage can be adapted to any other type of sensor (asymmetric pyramid for instance). Spatial filtering is a very simple, elegant and versatile solution (the size of the filter can be adapted to the observing conditions and the type of target). 

Future works will be dedicated to the optimisation of the filtering parameters and the full simulation in the realistic case of an ELT first generation instrument (HARMONI) 
Finally the spatially filtered non-modulated Fourier Filtered Wavefront Sensor (Pyramid, asymmetrical pyramid, other ...)  will be tested first on our laboratory bench LOOPS and then on sky using our newly integrated PAPYRUS bench at Observatoire de Haute Provence. 

\acknowledgments

This work benefited from the support of the WOLF project ANR-18-CE31-0018 of the French National Research Agency (ANR). It has also been prepared as part of the activities of OPTICON H2020 (2017-2020) Work Package 1 (Calibration and test tools for AO assisted E-ELT instruments). OPTICON is supported by the Horizon 2020 Framework Programme of  the  European  Commission’s  (Grant  number  730890). Authors are acknowledging the support by the Action Spécifique Haute Résolution Angulaire (ASHRA) of CNRS/INSU co-funded by CNES. Nicolas Levraud PhD is co-funded by  ONERA and INAF.

\bibliography{Bibliographie_SPIE} 
\bibliographystyle{spiebib} 

\end{document}